\documentclass[conference]{IEEEtran}
\usepackage{lineno,hyperref}
\hypersetup{
     colorlinks = true,
     linkcolor    = blue,
     citecolor    = red
}
\usepackage{bm}
\usepackage{mathrsfs}
\usepackage{amsmath,amsthm,amssymb,amsfonts}
\usepackage{graphicx}
\usepackage{subfigure}
\usepackage[dvipsnames]{xcolor}
\usepackage{algorithm,algorithmic}
\usepackage{cite}
\usepackage[nice]{nicefrac}
\usepackage{array}
\usepackage{balance}
\usepackage[flushleft]{threeparttable} 
\usepackage{enumerate}

\modulolinenumbers[5]

\makeatletter
\newcommand*{\transpose}{%
  {\mathpalette\@transpose{}}%
}
\newcommand*{\@transpose}[2]{%
  \raisebox{\depth}{$\m@th#1\intercal$}%
}
\newcommand{\thickhline}{%
    \noalign {\ifnum 0=`}\fi \hrule height 1pt
    \futurelet \reserved@a \@xhline
}
\newcolumntype{"}{@{\hskip\tabcolsep\vrule width 1pt\hskip\tabcolsep}}
\makeatother

\makeatletter
\g@addto@macro\normalsize{%
\setlength\abovedisplayskip{3pt}
\setlength\belowdisplayskip{3pt}
\setlength\abovedisplayshortskip{3pt}
\setlength\belowdisplayshortskip{3pt}
}
\makeatother

\usepackage{soul}
\usepackage{tikz}
\usepackage{booktabs}
\usepackage{tabularx}

\newcommand*\diff{\mathop{}\!\mathrm{d}}

\ifCLASSINFOpdf
\else
\fi

\begin{document}
\renewcommand{\ttdefault}{cmtt}
\bstctlcite{IEEEexample:BSTcontrol}

\title{Deep-Learning-Directed Preventive Dynamic Security Control via Coordinated Demand Response}

\author{\IEEEauthorblockN{Amin Masoumi and Mert Korkali}
\IEEEauthorblockA{\textit{Department of Electrical Engineering and Computer Science}\\ 
\textit{University of Missouri} \\
Columbia, MO 65211 USA \\
e-mail: \{\texttt{am4n5,korkalim\}@missouri.edu}}}

	

\markboth{2025 IEEE PES General Meeting, Austin, TX, USA}%
{Shell \MakeLowercase{\textit{et al.}}: Bare Demo of IEEEtran.cls for IEEE Journals}
\maketitle

\begin{abstract}

Unlike common faults, three-phase short-circuit faults in power systems pose significant challenges. These faults can lead to out-of-step (OOS) conditions and jeopardize the system's dynamic security. The rapid dynamics of these faults often exceed the time of protection actions, thus limiting the effectiveness of corrective schemes. This paper proposes an end-to-end deep-learning-based mechanism, namely, a convolutional neural network with an attention mechanism, to predict OOS conditions early and enhance the system's fault resilience. The results of the study demonstrate the effectiveness of the proposed algorithm in terms of early prediction and robustness against such faults in various operating conditions.

\end{abstract}
\IEEEpeerreviewmaketitle

\section{Introduction}
Dynamic security control (DSC) is essential in bulk power systems due to rare dynamic events (e.g., three-phase short-circuit faults).
Without proper DSC, these faults can lead to out-of-step (OOS) conditions \cite{OOS} or rotor angle instability, causing loss of synchronism among the synchronous generators and systemwide blackouts. The problem worsens during sudden load-level transitions or when the power system operates close to its stability margins (e.g., maximum loading). 
There are two approaches to mitigating the impact of dynamic events: corrective and preventive\cite{res} actions. Corrective approaches aim to localize and clear faults while activating adaptive maneuvers. These schemes can include voltage support, generator rescheduling, shedding, and dynamic reconfiguration. However, they are computationally complex and costly and may not be fully promising since transients occur on the order of milliseconds \cite{DSC} and propagate to other areas, causing maloperation.
Preventive approaches, on the other hand, are programmed based on prefault and fault data sets to respond effectively. 
Coordinated demand response (CDR) is a reliable scheme that targets affected areas promptly and adjusts demand to restore stability \cite{DR}. From a rational standpoint, this preemptive strategy curtails some portions of demand. 
In addition, the increased intensity of rotor angle instability can prevent it from spreading to other generators. In simpler terms, the preventive approach is cost-effective, meaning there is no need to install additional devices such as blackstart units. Moreover, it is promising to mitigate the spread of rare yet impactful events and restore stability. Therefore, it is imperative to prioritize the implementation of this preventive scheme. As previously mentioned, this approach relies on online monitoring. To achieve this, one must analyze the behavior of rare, three-phase short-circuit fault scenarios by evaluating transient instability status. More recently, deep learning (DL) models have shown promise in accurately and efficiently identifying fault patterns. These models can analyze extensive amounts of data and recognize patterns that indicate the potential for instability. By classifying contingency-impacted areas, operators can focus on specific regions and take appropriate measures to prevent blackouts. By intelligently utilizing CDR, operators can take timely and cost-effective preventive measures. Many of the existing DL models~\cite{Rt,R2, Sp-temp} leverage feature engineering to reduce the dimensionality of the input vector. While this approach works well for pure dynamic security assessment, it is disadvantageous for DSC because highly accurate assessment is the foundation for activating preventive approaches.

Employing a task-agnostic approach, this paper proposes a coordinated preventive control scheme based on time-domain simulation (TDS) to fully eliminate the destructive consequences of rare contingency events through a DL perspective. Therefore, the proposed solution methodology establishes a task-independent process, mitigating the occurrence of OOS conditions. Our paper makes the following key contributions:
\begin{enumerate}[(i)]
    \item addressing the prevention of transient instability by taking into account the uncertainty of three-phase short-circuit faults before entering the OOS condition; 
    \item proposing an end-to-end DL algorithm that captures the unstable behavior in short-circuit data sets; and
    \item evaluating the real-time performance of the DL-based algorithm in damping the impact of contingencies in resilience case studies using the proposed CDR.
\end{enumerate}

The remainder of this paper is structured as follows: Section~\ref{Sec.2} describes the studied scenarios and the proposed resilience metrics. Section~\ref{Sec.3} describes the proposed DLD-DSC scheme. Section~\ref{Sec.4} discusses the results and the performance validation. Lastly, Section~\ref{Sec.5} concludes the paper.

\vspace{-.2cm}
\section{Modeling Event Scenarios in TDS and the Proposed Resilience Metric}
\label{Sec.2}
The proposed DSC strategy is predicated on analyzing the impact of extreme dynamic event scenarios, particularly a three-phase short-circuit fault in Scenario $\upsilon$ ($\xi^{\text{fault}, x^\upsilon}$) using TDS. Hence, the status of transmission lines ($\xi^{\ell, \upsilon}$) is set to $0$ during the fault. Faults are simulated on all transmission lines varying the fault duration ($\tau^\upsilon$), the fault location ($x^\upsilon$), and system loading ($k^\upsilon$). They are selected based on uniform distributions in the ranges of $\tau^\upsilon=[0.06, 0.4]$ s, $x^\upsilon=[0, 100] \%$ and $k^\upsilon=[0.75,1.5]$ \rm{pu}. 
Faults are applied at $t_\text{start}=2$ s, and the impact is monitored for a duration of $5$ s and $7$ s to be compatible with standard environments \cite{Rt}. For this study, we select the IEEE $39$-bus ($10$-machine New England) system as the test case due to its capability of capturing the oscillation in rotor angles of bulk power systems. 
In addition, it includes critical lines and areas of concentrated power flow, which are suitable for studying cascading failures and evaluating the value of preventive approaches. Hence, TDS is conducted after setting the inputs, fault duration, fault location, and system loading to assess the impact of rare event scenarios. Finally, the results are recorded to provide a clear insight into their impact by deriving the resilience metric\cite{res}, $R = \int_{{t_{\text{start}}}}^{{t_{\text{end}}}} {{S_{{\text{rms}}}}\diff t} $ and so the normalized value $\widehat{R} = \int_{{t_{{\text{start}}}}}^{{t_{{\text{end}}}}} {k \diff t} = k(t_\text{end}-t_\text{start})=k\tau$, where $t_{\text{start}}$, $t_{\text{end}}$, and $S_{{\text{rms}}}$ denote the fault start time, fault end time, and normalized root-mean-square (rms) value of composite loads, respectively. The full description of the modeling is presented in Algorithm~\ref{alg1}, where $\xi^{\ell, \upsilon}$ and $\xi^{\text{fault}, x^\upsilon}$ are the status of Line $\ell$ and the status of the fault based on its location ($x^\upsilon$) on Line $\ell$, respectively. 
\begin{algorithm}
 \caption{Modeling Three-Phase Short-Circuit Scenarios in a TDS Environment}
 \label{alg1}
 \begin{algorithmic}[1]
 \renewcommand{\algorithmicrequire}{\textbf{Input:}}
 \renewcommand{\algorithmicensure}{\textbf{Output:}}
 \REQUIRE $\tau^\upsilon; k^\upsilon; x^\upsilon$; $\xi^{\ell, \upsilon}, \ \ell=\{1,\dots,46\}; \xi^{\text{fault}, x^\upsilon}$
 \ENSURE $R^\upsilon$
 \\ \textit{Initialization}: $t_{\text{start}}^{\ell,\upsilon} = t_\text{start}, t_{\text{end}}^{\ell,\upsilon} = t_\text{start} + \tau^\upsilon, S_{{\text{rms}}}^\upsilon = k^\upsilon S_{{\text{rms}}}, \rm{timevector}=0:7$ s
 \FOR {$\upsilon = 1:V$}
    \STATE $S_{{\text{rms}}}^\upsilon = k^\upsilon S_{{\text{rms}}}$
    \FOR {$t \in \rm{timevector}$}
        \IF {($t \geq t_{\text{start}}^{\ell,\upsilon}$ \textbf{and} $t \leq t_{\text{end}}^{\ell,\upsilon}$)}
            \STATE $\xi^{\ell, \upsilon} = 0, \xi^{\text{fault}, x^\upsilon} = 1$
        \ELSE
            \STATE $\xi^{\ell, \upsilon} = 1, \xi^{\text{fault}, x^\upsilon} = 0$
        \ENDIF
    \ENDFOR
    \RETURN $R^\upsilon$
 \ENDFOR
 \end{algorithmic} 
\end{algorithm}
We employ two metrics to quantify the contribution of the DLD-DSC by performing CDR. In this regard, we define the \emph{system failure index} (SFI) as the probability of complete system collapse after reaching \emph{average system resilience} (ASR) under unstable scenarios, fault uncertainty, location, duration, and system loading. The expected result is the conversion of the $k$ to $k^\text{CDR}$. The SFI is formulated as follows:
\vspace{-0.2cm}
\begin{subequations}
\label{eq.1}
\begin{align}
\text{SFI} & \Rightarrow F_{\widehat{R}}(r) \\
& =P(k \tau \leqslant r) \\
& =P(k \leqslant r / \tau \mid \tau>0) P(\tau>0) \nonumber \\
& \qquad + P(k \geqslant r / \tau \mid \tau<0) P(\tau<0) \\
& =P(k \leqslant r / \tau, \tau>0)+P(k \geqslant r / \tau, \tau<0) \\
& =\int_0^{\infty} \int_{-\infty}^{r / \tau} f_{k \tau}(k, \tau) \diff k \diff \tau \nonumber\\
& \qquad + \int_{-\infty}^0 \int_{r / \tau}^{\infty} f_{k \tau}(k, \tau) \diff k \diff \tau \\
f_{\widehat{R}}(r) & =\frac{\partial F_R(r)}{\partial r} \\
& =\int_0^{\infty} \frac{1}{\tau} f_{k \tau}\left(\frac{r}{\tau}, \tau\right) \diff \tau \nonumber\\
& \qquad +\int_{-\infty}^0 \frac{-1}{\tau} f_{k \tau}\left(\frac{r}{\tau}, \tau\right) \diff \tau \\
& =\int_{-\infty}^{\infty} \frac{1}{|\tau|} f_{k \tau}\left(\frac{r}{\tau}, \tau\right) \diff \tau,\\
& \qquad k^{\mathrm{CDR}} \in[\alpha, \beta], \ \tau \in[0.06,0.4] \nonumber
\end{align}
\end{subequations}
\vspace{-0.2cm}

Now, we determine the system failure probability based on ASR and $k^\text{CDR}$ since $k$ and $\tau$ are independent variables, i.e., ${f_{k\tau }}(\frac{r}{\tau },\tau ) = {f_k}(\frac{r}{\tau }){f_\tau }(\tau )$. Also, we know that 
\begin{subequations}
\small
\label{eq.2}
\begin{align}
\!\!\!\!\!
f_k\left(\frac{\text{ASR}}{\tau}\right)&= \left\{\begin{array}{c}
\!\!\!\!\!\!\!\!\!\!\!\!
\frac{1}{(\beta-\alpha)} \text { if } \alpha \leqslant \frac{\text { ASR }}{\tau} \leqslant \beta \\
\qquad \Rightarrow\frac{\text { ASR }}{\beta} \leqslant \tau \leqslant \frac{\text { ASR }}{\alpha}\\
0 \quad \text { otherwise }
\end{array}\right. \\
f_\tau(\tau)&=\left\{\begin{array}{c}
\!\!\!
\frac{1}{(0.4-0.06)} \text { if } 0.06 \leqslant \tau \leqslant 0.4 \\
0 \quad \text { otherwise }
\end{array}\right. \\
f_{\widehat{R}}(\text{ASR})&=\int_{-\infty}^{\infty} \frac{1}{|\tau|} f_{k \tau}\left(\frac{\mathrm{ASR}}{\tau}, \tau\right) \diff \tau \\
& =\int_{-\infty}^{\infty} \frac{1}{|\tau|} f_k\left(\frac{\text{ASR}}{\tau}\right) f_\tau(\tau) \diff \tau\\
\Rightarrow \text{SFI}=f_{\widehat{R}}(\text{ASR})&=\frac{1}{0.34(\beta-\alpha)} \ln \left(\frac{\min \left\{0.4, \frac{\mathrm{ASR}}{\alpha}\right\}}{\max \left\{0.06, \frac{\text{ASR}}{\beta}\right\}}\right)
\end{align}  
\end{subequations}
\begin{figure*}[th!]
\centering
\includegraphics[width=1.34\textwidth]{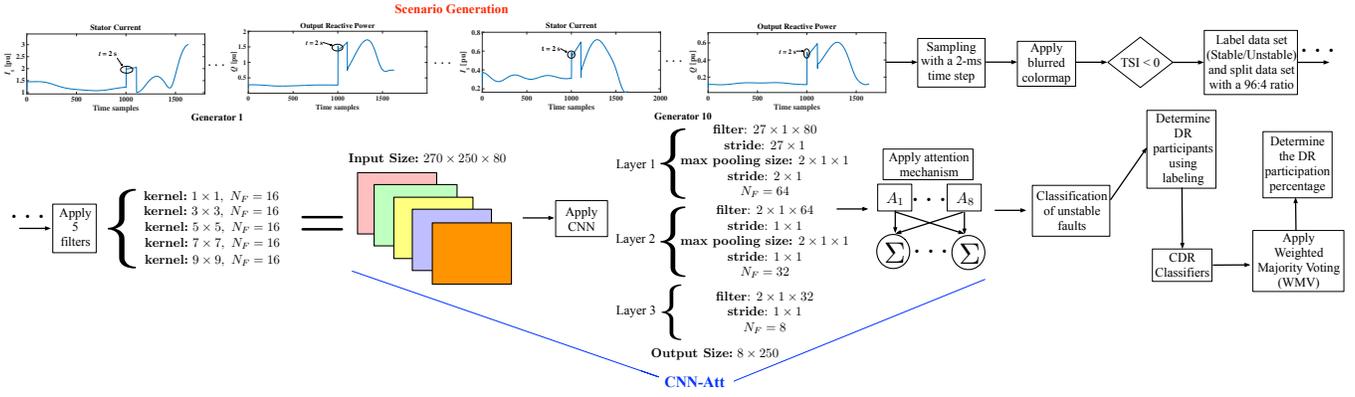}
 \vspace{-0.6cm}
\caption{The proposed DLD-DSC via CDR.}
\label{fig:Fig1}
\vspace{-0.3cm}
\end{figure*}

\vspace{-0.3cm}
\section{DL-Directed DSC (DLD-DSC) via CDR}
\label{Sec.3}
This section explains the proposed procedure for obtaining resilient DLD-DSC. The first task is to determine rotor angle instability by studying three-phase short-circuit faults. The second task involves coordinating the demand response (DR) strategy with critical areas and preventing the OOS condition, as discussed in the following subsections.

\subsection{TDS Environment and Transient Instability Status}

In this section, we will discuss the TDS environment and instability criteria. For offline training, the proposed DL-based algorithm focuses on rare event scenarios in TDS. We assume that measurements are taken at each generator bus to evaluate the impact on generator dynamics (i.e., excitation and governor). The sampling frequency is set to $2$ ms. The offline training utilizes prefault and fault data sets to classify the OOS condition. This is important for online monitoring and testing the effectiveness of classifying the OOS condition. 
A sliding window with a length of $0.5$ s is used to capture features for instability classification. This window captures prefault data ranging from $0.44$ s to $0.1$ s before fault clearance. The study considers $27$ features, divided into $9$ primary (i.e., instantaneous measurements ~\cite{Rt} [i.e., $i_d$, $i_q$, $v_d$, $v_q$, $\delta$, $\omega_m$, $T_e$, $P_g$, and $Q_g$]), $9$ secondary, and $9$ tertiary sets. The second set of features is derived by subtracting consecutive samples from the primary set. The last set is obtained based on the difference between rotor angles of generators (i.e., $\delta_{i,j} = |\delta_{i}-\delta_{j}|$). The \emph{transient instability status} (TIS)\cite{TIS} is calculated after observing the impact of rare event scenarios on the rotor angles during the $7$-s TDS as follows:
\begin{equation}
\label{eq.3}
{\text{TIS = }}\left\{ \begin{gathered}
  \text{class 1}{\text{, if}} \ \frac{{360 - {\Lambda ^{\max }}}}{{360 + {\Lambda ^{\max }}}} < 0 \hfill \\
  {\text{class 0, if}} \ \frac{{360 - {\Lambda ^{\max }}}}{{360 + {\Lambda ^{\max }}}} > 0 \hfill \\ 
\end{gathered}  \right.
\end{equation}

\noindent where ${\Lambda ^{\max }} = \max \{|{\delta_i} - {\delta_j}|\},  \ i, j \in \{1,\dots,10\}$, is the maximum difference between angles of any two generators.
 
  

\vspace{-.2cm}
\subsection{Proposed Methodology}

\begin{figure}[h!]
\centering
\includegraphics[width=.87\columnwidth,keepaspectratio=true,clip=true]{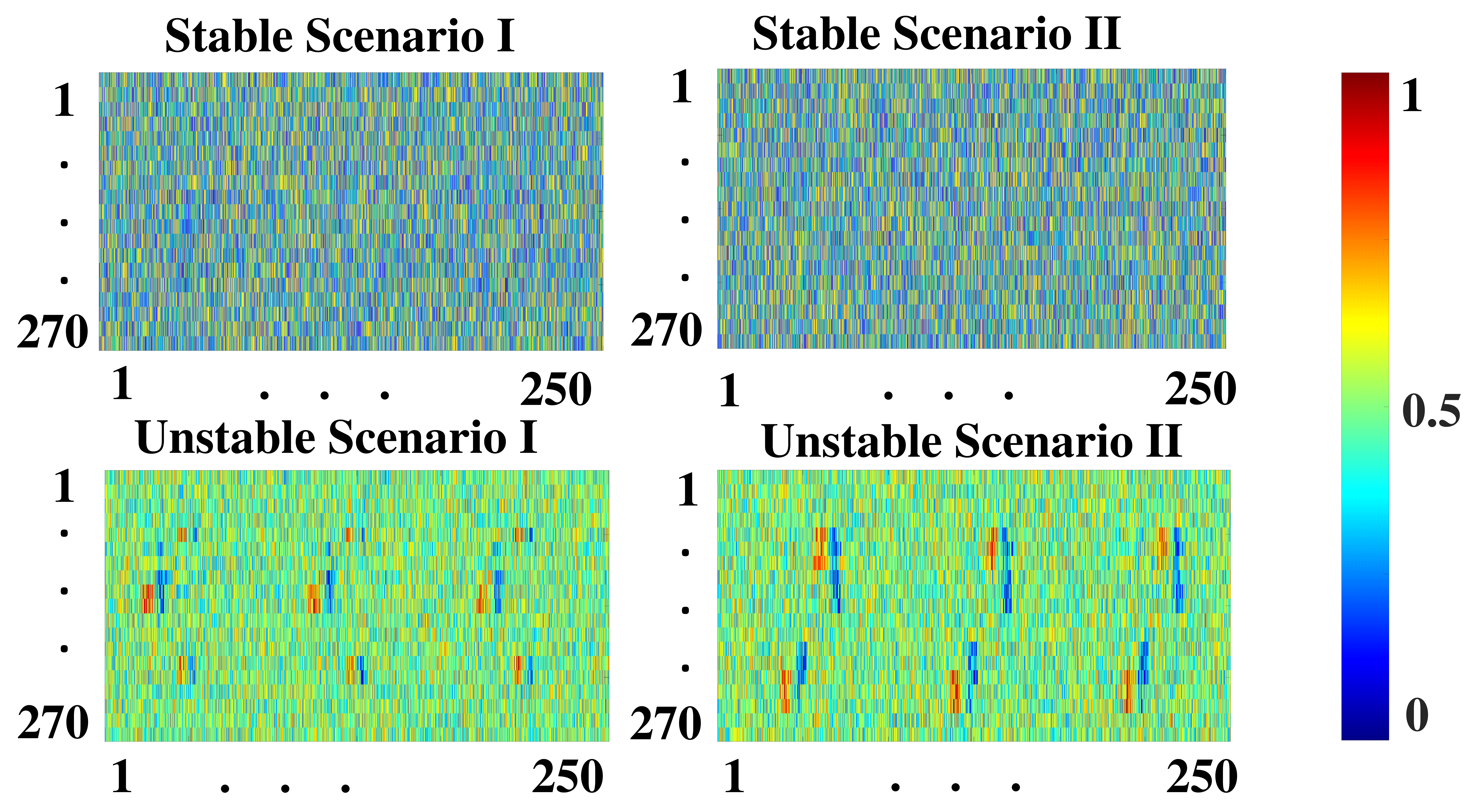}
 \vspace{-0.4cm}
\caption{The blurred colormap representation of the tabular data.}
\vspace{-0.5cm}
\label{fig:Fig2}
\end{figure}

We perform TDS on the IEEE $39$-bus system, with $27$ features each, and record the results of prefault and fault states, sampling every $2$ ms for $0.5$ s, with $250$ time samples. Hence, the created tabular data size is $270 \times 250$. Fig.~\ref{fig:Fig1} shows a sudden transition before and after the contingency. 
Here, we propose using the colormap intensity representation~\cite{R2} to differentiate features before and after contingency. To do this, we calculate the intensity of the blurred colormap applied on tabular data, as shown in Fig.~\ref{fig:Fig2}. Two pairs of stable and unstable contingencies based on the TIS are also presented. Fig.~\ref{fig:Fig2} shows that unstable pairs can be distinguished from stable pairs based on colormap intensity, specifically $0$-$0.5$ and $0.5$-$1$, respectively. In unstable scenarios, distinct dense vertical areas indicate the spatiotemporal impact \cite{Sp-temp} of dynamic contingencies on power system oscillations. 

Certain features exhibit abnormal behavior during a fault, specifically in the blue and red areas. This behavior is primarily influenced by the dynamics of the IEEE $39$-bus system and the sensitivity of these features to faults, resulting in spatial entanglement. The recurrence of this spatial entanglement in TDS indicates temporal dependence during fault clearance. 

For the classification problem at hand, we selected a convolutional neural network (CNN) as the most viable approach. However, the challenge lies in decoding the spatiotemporal impact by establishing a connection between their underlying behavior, which is common to all scenarios. We assume that the vertical and narrow anomalies, represented as blurred colormap intensities across $270$ features, are consistently observed in these two distinct unstable scenarios. Thus, amplifying the intensity in these areas can enhance the classification quality since they share the same visual characteristics. We propose an algorithm synthesizing a CNN algorithm combined with an attention mechanism\cite{attention,wmv} (denoted CNN-Att henceforth) and apply multiple filters to the tabular data. This allows us to uncover the interdependence of spatial deep features shared among the generators. The volume data, which is the tabular data concatenated by $5$ filters, is inputted into the CNN to decode the significance of spatial data. Another challenge that must be addressed during contingencies is the temporal interdependency of hidden features. We use the attention mechanism to determine the correlation between prefault and fault time samples and feed them into a dense feedforward neural network along with the softmax operator to perform the classification of TIS. We aim to design a preventive strategy called CDR to counter OOS conditions. We conduct an $N-1$ dynamic contingency analysis on $46$ transmission lines to identify the critical loads. This analysis is performed under maximum loading and the maximum duration of a three-phase short-circuit fault. We observe TIS when an OOS condition occurs. This helps us identify the $30$ most critical lines. Next, we identify the critical loads directly connected to these $30$ critical lines. We calculate the ASR, which is $0.48$. This value is obtained through $10{,}000$ multi-level Monte Carlo (MLMC) simulations for the $46$ lines, where we vary the fault location, duration, and system loading. We implement a load-shedding strategy for the critical loads using the ASR as a threshold. This strategy is carried out hierarchically based on the maximum system allowance to participate in DR. The maximum allowance ranges from $1\%$ to $10\%$ in steps of $1\%$. The purpose of this strategy is to redistribute the demand to local generators if they fail to supply the critical loads. 
If an unstable scenario's impact on the resilience exceeds the ASR, noncritical loads connected to noncritical lines are engaged in DR. This allows us to coordinate the critical and noncritical loads as the first and second participants of DR, respectively. This coordination is achieved through labeling as follows:
\begin{equation}
\label{eq.4}
{\text{Label =  }}\left\{ \begin{gathered}
  {\text{class 1    ,   if  }}{{\widehat{R}}^\upsilon } \leqslant {\text{ASR}} \hfill \\
  {\text{class 0    ,   otherwise}} \hfill \\ 
\end{gathered}  \right.
\end{equation}
\begin{subequations}
 \label{eq.5}
\begin{align}
\mathrm{DR}^\upsilon & =\left\{\begin{array}{l}
X_1^\upsilon \mathrm{DR}_1^\upsilon \text { if Label }=\text { class } 1 \\
X_2^\upsilon \mathrm{DR}_2^\upsilon \text { if Label }=\text { class } 0
\end{array}\right. \\ \label{eq.CDR}
\mathrm{CDR} & =\frac{\sum\limits_\upsilon^V \mathrm{DR}^\upsilon}{V} \\ 
X_1^v & + X_2^\upsilon=1, \ X_1^\upsilon, X_2^\upsilon \in\{0,1\} \ \ \forall \upsilon
\end{align}
\end{subequations}
\noindent where Classes 1 and 0 belong to critical and noncritical loads, respectively. This equation models binary decision variables in the implementation of CDR. ${\text{DR}}_1^\upsilon$ and ${\text{DR}}_2^\upsilon$ are the participation percentages of critical and noncritical loads, respectively. To implement the proposed preventive approach, we utilize the transfer-learning-based weighted majority voting (TL-WMV) algorithm to activate the CDR. TL-WMV leverages the pattern identification capabilities of the reliable DL algorithm. We fine-tune the most reliable DL algorithm using the TL approach~\cite{TL} as they share a common input vector. Subsequently, WMV takes charge of the second task by employing the labeling approach. WMV is a machine-learning (ML) technique that combines the knowledge from multiple classifiers by weighing their predictions and consequently enhancing the accuracy of classifying DR participants~\cite{wmv}. It is important to note that we determine the weight of each classifier's prediction based on its precision, which is calculated by dividing the classifier's precision by the sum of all classifier precisions\cite{wmv}. 
This approach ensures that targeted DR participants are activated, preventing the OOS condition. After performing the TL-WMV algorithm, a dense feedforward neural network is trained to estimate the participation percentage of critical and noncritical loads (i.e., ${\text{DR}}_1^\upsilon$ and ${\text{DR}}_2^\upsilon$).


\section{Results and Discussion}
\label{Sec.4}
We utilized MATLAB\textsuperscript{\textregistered} R$2024$b and TensorFlow platform (Python 3.8) for performing TDS and running the proposed DL-based algorithm, respectively. Fault resistance, as well as the first- and second-shot duration of the autoreclosing scheme, are set to $0.001~\Omega$, $20$ ms, and $5$ s, respectively. Moreover, $10$-fold cross-validation is applied for the data size of 4000 scenarios. The configuration details of the CNN-Att, including kernel/filter size, number of filters ($N_F$), padding (same for kernels and none for filters), stride size, and attention size, are listed in Fig.~\ref{fig:Fig1}. In addition, the maximum (max) pooling size is $2 \times 1 \times 1$. For classification, we used a learning rate of $0.0001$, a batch size of $64$, a dropout of $0.2$, and the Adam optimizer. The regression task involves $5$ hidden layers with $300$/$300$/$250$/$250$/$200$ neurons. 

\begin{figure}[ht!]
\centering
\includegraphics[width=0.38\textwidth]{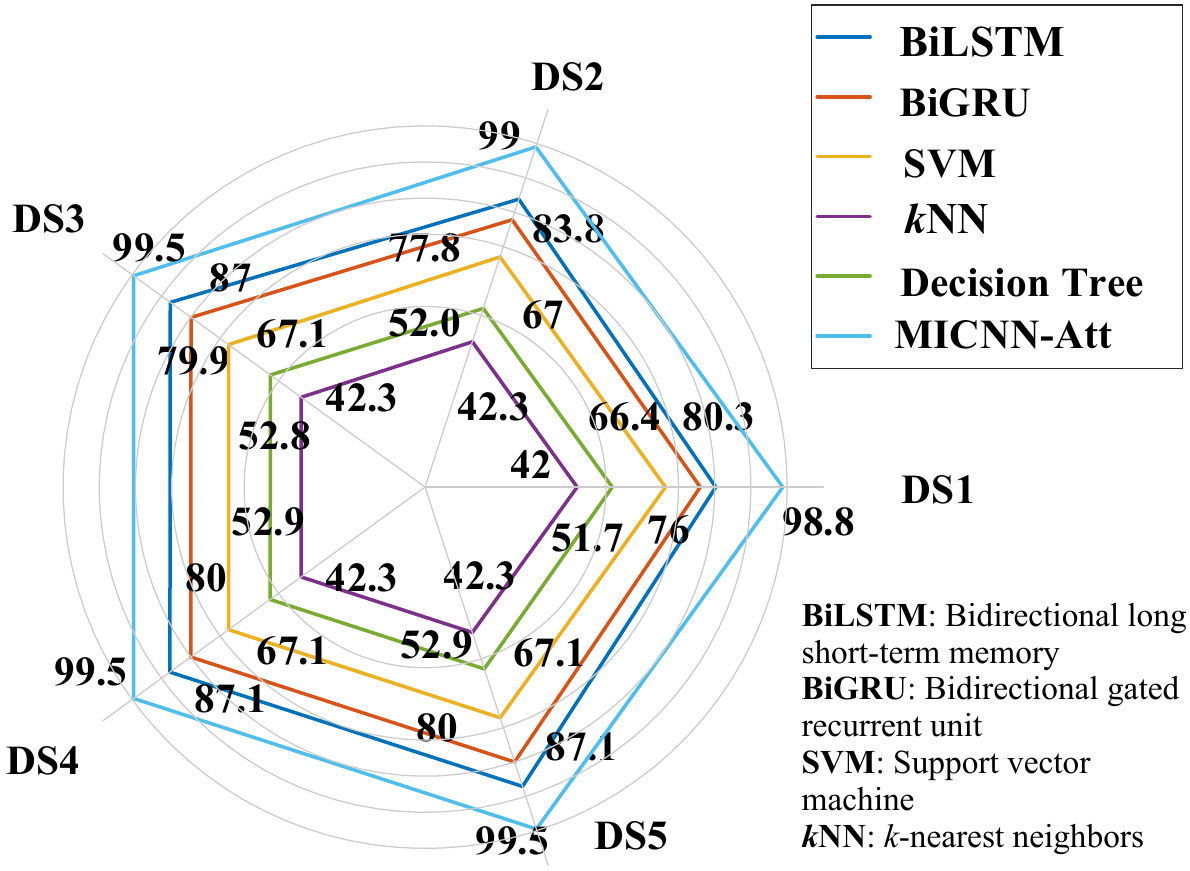}
 \vspace{-0.4cm}
\caption{The comparison of CNN-Att accuracy in rotor angle instability identification.}
\label{fig:Fig3}
\end{figure}

To demonstrate the effectiveness of the CNN-Att algorithm, we classified the rotor angle instability of $5$ volumes of the data sets (DSs), i.e., $\rm{DS}1=2{,}000$, $\rm{DS}2=3{,}000$, $\rm{DS}3=4{,}000$, $\rm{DS}4=5{,}000$, and $\rm{DS}5=6{,}000$. As shown in Fig.~\ref{fig:Fig3}, CNN-Att achieved higher accuracy than the ML- and DL-based techniques for all DSs. Notably, we achieved a higher accuracy of $98.8\%$ for the lower volume of DS ($2{,}000$ scenarios), which demonstrates the effectiveness of the proposed CNN-Att model in distinguishing between stable and unstable scenarios.
\vspace{-0.6cm}

\begin{table}[h!]
\vspace{-.7cm}
\caption{Accuracy of Classification in Performing CDR}
\vspace{-.5cm}
\label{Table1}
\begin{center}
\begin{tabular}{c|c}
\hline
\hline
\textbf{CDR classifiers} & \textbf{Accuracy} [$\%$]  \\
\hline
BiLSTM ($\rm{DS}2$) & $88.81\%$  \\

BiLSTM ($\rm{DS}3$) & $92.37\%$ \\
BiLSTM ($\rm{DS}4$) & $95.62\%$ \\
BiLSTM ($\rm{DS}5$) & $96\%$ \\
BiGRU ($\rm{DS}2$) & $86.93\%$ \\
BiGRU ($\rm{DS}3$) & $88.48\%$\\
BiGRU ($\rm{DS}4$)  & $92.11\%$  \\

BiGRU ($\rm{DS}5$) & $93.04\%$ \\

SVM with Gaussian kernels ($\rm{DS}2$) & $70.79\%$ \\
SVM with Gaussian kernels ($\rm{DS}3$) & $71.03\%$ \\
SVM with Gaussian kernels ($\rm{DS}4$) & $71.9\%$ \\

SVM with Gaussian kernels ($\rm{DS}5$) & $73.15\%$ \\
CNN-Att ($\rm{DS}3$) & $99.1\%$\\
\textbf{TL-WMV} & {$\bf{99.7}\%$} \\
\hline
\hline
\end{tabular}
\end{center}
\vspace{-.5cm}
\end{table}

Table~\ref{Table1} shows the dependability of TL-WMV based on the prediction results. We used a pretrained set of the DL-based algorithm to predict the best match for participation in the proposed CDR. Precautions were taken to exclude the initial variants of all pretrained networks to classify noncritical and critical loads. The same logic was applied to the support vector machine (SVM), which performed rather poorly.

Table~\ref{Table2} demonstrates the reliability of the proposed DL approach compared to all ML-based algorithms, including SVM, Gaussian process regression (GPR), and decision trees (DT). We measured the participation percentage of actors (i.e., ${\text{DR}}_1^\upsilon$ and ${\text{DR}}_2^\upsilon$ in CDR) and obtained the root-mean-square error (RMSE).
\vspace{-.3cm}
\begin{table}[h!]
\caption{Estimation of Demand Response Performance}
\vspace{-.2cm}
\label{Table2}
\begin{center}
\begin{tabular}{c|c|c|c|c}
\hline
\hline
\textbf{Regressors} & SVM & GPR & DT &\textbf{Proposed}\\
\hline

\textbf{RMSE} & $0.157$ & $0.083$ & $0.2$ & $\mathbf{0.0041}$\\
\hline
\hline
\end{tabular}
\end{center}
\vspace{-.3cm}
\end{table}

Table~\ref{Table3} shows the explainable features from the blurred colormap representation. It provides each filter's average normalized weight (ANW).  
\begin{table}[h!]
\caption{The Performance of Kernels for Encoding the Deep Spatial Information}
\vspace{-.4cm}
\label{Table3}
\begin{center}
\begin{tabular}{c|c|c|c|c|c}
\hline
\hline
\textbf{Kernel size}& $1\times1$ & $3\times3$ & $5\times5$ &$7\times7$&$9\times9$\\
\hline

\textbf{ANW} & $0.04$ & $0.1$ & $\mathbf{0.4}$ & $\mathbf{0.4}$ & $0.06$\\
\hline
\hline
\end{tabular}
\end{center}
\vspace{-.2cm}
\end{table}
Consequently, $5\times 5$ and $7\times 7$ kernels were assigned at a higher frequency to the blurred colormap data. As shown in Fig.~\ref{fig:Fig2}, most of the features with high oscillation in unstable case studies belong to the lower parts of the blurred colormap, which includes Generators $6$, $7$, $8$, $9$, and $10$. Among these, Generators $6$, $7$, and $9$ are directly connected to $30$ critical lines. This confirms that the proposed algorithm successfully captures a specific contingency pattern. Another concern is decoding the temporal dependency. The ANW of the hidden features of each time sample in the attention layer is calculated. Based on Table~\ref{Table4}, two hidden features with specific indices have captured the importance weight. 
\begin{table}[h!]
\vspace{-.2cm}
\caption{The Index Pairs of the Shared Temporal Features in the Attention Blocks}
\vspace{-.4cm}
\label{Table4}
\begin{center}
\begin{tabular}{c|c|c|c|c}
\hline
\hline
& $A_1$ & $A_2$ & $A_3$ &$A_4$ \\
\hline
\textbf{Indices} & $46$ and $249$ & $40$ and $248$ & $43$ and $246$ & $48$ and $248$\\
\hline
\hline
 &$A_5$ &$A_6$ &$A_7$ & $A_8$\\
\hline
\textbf{Indices} & $41$ and $247$ & $43$ and $245$ & $47$ and $248$& $44$ and $249$\\
\hline
\hline
\end{tabular}
\end{center}
\vspace{-.2cm}
\end{table}
These hidden features, which are the most shared information across all attention blocks, i.e., $\{A_1,\dots,A_8\}$, belong to the first and last sets of hidden features of time samples. This is because of the difficulty in detecting the worst-case scenarios of stable and unstable faults, particularly faults with a duration of $0.06$ s that do not cause OOS and faults with a duration of $0.4$ s that lead to OOS. 
Faults begin at $t=2$ s and take $0.06$ s and $0.4$ s to clear for stable and unstable scenarios, respectively. For the stable scenario, there are only $30$ fault and $220$ prefault time samples available. The attention module prioritizes fault time samples as they help differentiate between fault scenarios with similar durations but with unstable (i.e., OOS) conditions. In the unstable scenario, $50$ prefault and $200$ fault time samples are accessible. It is logical to focus on 
the $50$ prefault time samples to distinguish from the same fault duration ($0.4$) but with stable behavior (not leading to OOS). Both the stable and unstable scenarios have $30$ and $50$ time samples, respectively, in the instantaneous zone of distance relay (Zone 1). The attention layer aligns with the tripping pattern of the distance relay, which controls the recloser switches and affects the oscillation of spatial features displayed in the blurred colormap. The proposed algorithm effectively links hidden features consistently across prefault and fault data. The default value of SFI, based on the $N-1$ contingency, is $0.87$, with an $\text{ASR}$ of $0.48$. This means that in the worst-case scenario of a three-phase short-circuit fault with a maximum duration and maximum loading, there is a probability of $0.87$ that the system's resilience will be lower than $0.48$. We assess the SFI based on the obtained CDR of $5\%$ load shedding using (\ref{eq.1}) and (\ref{eq.2}).


The SFI was found to be $0.71$ using CDR (\ref{eq.CDR}). In other words, if we practice DR and drive $k^\text{CDR}$ to $[0.7125, 1.425]$, we will suppress the probability of power system collapse from $0.87$ to $0.71$, enhancing the robustness against unstable faults.

\section{Conclusions}
This paper proposes a combination of CNN and an attention mechanism that leverages end-to-end deep learning to mitigate three-phase faults leading to instability. The proposed algorithm offers a task-agnostic solution methodology with a focus on early prediction and coordinated activation of DR, using both classification and regression techniques. The results of our study show that the algorithm improves robustness against such faults. Future research directions involve applying the algorithm to real-world case studies through optimal corrective schemes for complete fault recovery.
\label{Sec.5}



\ifCLASSOPTIONcaptionsoff
\newpage
\fi

\bibliographystyle{IEEEtran}
\bibliography{IEEEabrv,References}

\begin{thebibliography}{10}
\providecommand{\url}[1]{#1}
\csname url@samestyle\endcsname
\providecommand{\newblock}{\relax}
\providecommand{\bibinfo}[2]{#2}
\providecommand{\BIBentrySTDinterwordspacing}{\spaceskip=0pt\relax}
\providecommand{\BIBentryALTinterwordstretchfactor}{4}
\providecommand{\BIBentryALTinterwordspacing}{\spaceskip=\fontdimen2\font plus
\BIBentryALTinterwordstretchfactor\fontdimen3\font minus
  \fontdimen4\font\relax}
\providecommand{\BIBforeignlanguage}[2]{{%
\expandafter\ifx\csname l@#1\endcsname\relax
\typeout{** WARNING: IEEEtran.bst: No hyphenation pattern has been}%
\typeout{** loaded for the language `#1'. Using the pattern for}%
\typeout{** the default language instead.}%
\else
\language=\csname l@#1\endcsname
\fi
#2}}
\providecommand{\BIBdecl}{\relax}
\BIBdecl

\bibitem{OOS}
M.~A. Aftab, M.~M. Roomi, C.~Konstantinou, and S.~S. Hussain, ``Out-of-step
  protection in power system,'' in \emph{Power System Protection in Future
  Smart Grids}.\hskip 1em plus 0.5em minus 0.4em\relax Elsevier, 2024, pp.
  31--58.

\bibitem{res}
N.~Bhusal, M.~Abdelmalak, M.~Kamruzzaman, and M.~Benidris, ``Power system
  resilience: Current practices, challenges, and future directions,''
  \emph{IEEE Access}, vol.~8, pp. 18\,064--18\,086, 2020.

\bibitem{DSC}
L.~Duchesne, E.~Karangelos, and L.~Wehenkel, ``Recent developments in machine
  learning for energy systems reliability management,'' \emph{Proc. {IEEE}},
  vol. 108, no.~9, pp. 1656--1676, Sep. 2020.

\bibitem{DR}
M.~R.~V. Moghadam, R.~T. Ma, and R.~Zhang, ``Distributed frequency control in
  smart grids via randomized demand response,'' \emph{{IEEE} Trans. Smart
  Grid}, vol.~5, no.~6, pp. 2798--2809, Nov. 2014.

\bibitem{Rt}
F.~Darbandi, A.~Jafari, H.~Karimipour, A.~Dehghantanha, F.~Derakhshan, and
  K.-K. Raymond~Choo, ``Real-time stability assessment in smart cyber-physical
  grids: A deep learning approach,'' \emph{IET Smart Grid}, vol.~3, no.~4, pp.
  454--461, Aug. 2020.

\bibitem{R2}
A.~Gupta, G.~Gurrala, and P.~Sastry, ``An online power system stability
  monitoring system using convolutional neural networks,'' \emph{{IEEE} Trans.
  Power Syst.}, vol.~34, no.~2, pp. 864--872, Mar. 2019.

\bibitem{Sp-temp}
M.~Yu, J.~Sun, S.~Tian, S.~Zhang, J.~Wei, and Y.~Wu, ``Identification of
  dominant instability modes in power systems based on spatial-temporal feature
  mining and {TSOA} optimization,'' \emph{IET Gener. Transm. Distrib.}, 2024.

\bibitem{TIS}
A.~Iqbal and T.~Jain, ``Phasor area criterion--wide area assessment of
  transient stability,'' \emph{{IEEE} Trans. Power Syst.}, vol.~39, no.~6, pp.
  6876--6888, Nov. 2024.

\bibitem{attention}
A.~Rolander, A.~Ter~Vehn, R.~Eriksson, and L.~Nordstr{\"o}m, ``Real-time
  transient stability early warning system using {G}raph {A}ttention
  {N}etworks,'' \emph{Electr. Power Syst. Res.}, vol. 235, p. 110786, Oct.
  2024.

\bibitem{wmv}
A.~Masoumi and M.~Korkali, ``Deep-learning-enhanced static risk-oriented
  security assessment under uncertainty,'' in \emph{Proc. IEEE Power \& Energy
  Soc. Gen. Meet.}, 2024, pp. 1--5.

\bibitem{TL}
C.~Ren and Y.~Xu, ``Transfer learning-based power system online dynamic
  security assessment: Using one model to assess many unlearned faults,''
  \emph{{IEEE} Trans. Power Syst.}, vol.~35, no.~1, pp. 821--824, Jan. 2020.

\end{thebibliography}
	
\end{document}